# A Survey of Data Compression Algorithms and their Applications

Mohammad Hosseini
Network Systems Lab
School of Computing Science
Simon Fraser University, BC, Canada
Email: mohammad_hosseini@sfu.ca

*Abstract*—**Today, with the growing demands of information storage and data transfer, data compression is becoming increasingly important. Data Compression is a technique which is used to decrease the size of data. This is very useful when some huge files have to be transferred over networks or being stored on a data storage device and the size is more than the capacity of the data storage or would consume so much bandwidth for transmission in a network. With the advent of the Internet and mobile devices with limited resources, data compression has gained even more importance. It can be effectively used to save both storage and bandwidth, thus to decrease download duration. Data compression can be achieved by a host of techniques. During this survey, I'm going to thoroughly discuss some of important data compression algorithms, their performance evaluation, and their major applications along with today's issues and recent research approaches.**

## I. Introduction

Data compression is one of the enabling technologies for multimedia applications. It would not be practical to put images, audio and video on websites if do not use data compression algorithms. Mobile phones would not be able to provide communication clearly without data compression. With data compression techniques, we can reduce the consumption of resources, such as hard disk space or transmission bandwidth. In this survey, first we introduce the concept of lossy and lossless data compression techniques, and will thoroughly discuss some of their major algorithms. Then we pick two of the mostly used compression algorithms, implement them and compare their compression ratio as a performance factor. Then in the third part, we discuss some of the most significant applications of data compression algorithms in multimedia compression, JPEG and MPEG coding algorithms. Finally in the last section we would enumerate some recent issues and approaches regarding data compression, like energy consumption.

## II. Data Compression Algorithms: Lossy and Lossless compression

Basically, we have two types of data compression algorithms. Lossless algorithms, which can reconstruct the original message exactly from the compressed message, and lossy algorithms, which can only reconstruct an approximation of the original message. Lossless algorithms are typically used for text, and lossy compression algorithms, on the other hand, remove unnecessary data permanently and so the original data cannot be completely regenerated. This is used for image, sound and video compression since it can cause significant reduction in file size with no significant quality reduction.

## III. Lossless Data Compression

### A. *Huffman Algorithm*

Huffman coding algorithm was first developed by David Huffman as part of a class assignment! The class was the first ever in the area of information theory and was taught by Robert Fano at MIT. The codes generated by this algorithm are called Huffman codes. These codes are prefix codes and are optimum for a given model (set of probabilities). The Huffman procedure is based on two observations regarding optimum prefix codes:

1) In an optimum code, symbols that occur more frequently (have a higher probability of occurrence) will have shorter codewords than symbols that occur less frequently.
2) In an optimum code, the two symbols that occur least frequently will have the same length.

It is easy to see that the first observation is correct. If symbols that occur more often had codewords that were longer than the codewords for symbols with less number of occurrences, the average number of bits per symbol would be larger comparing with the reverse case. Therefore, a code that assigns longer codewords to symbols that occur more frequently cannot be optimum. [1]

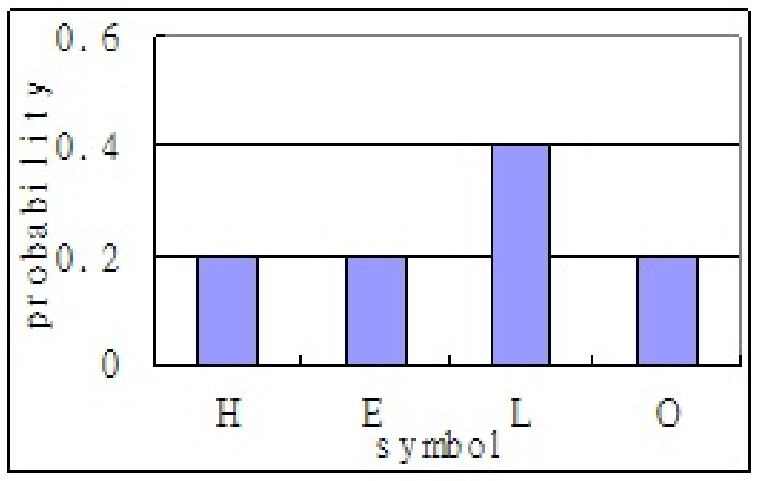

Fig. 1. Probabilities of alphabets in the example word

Huffman coding is used as a variable length coding. Characters in files are stored in ASCII codes, each taking up exactly 1 byte. This is a fixed-length code. The Huffman algorithm assigns codes to the characters depending on the number of repentance occur in the file which gives the shortest code to the most frequently occurring character. This results in a variable length code. Huffman codes are uniquely decodable.

*Algorithm:* The algorithm is as follows:

1) Select two symbols with the least probability form a binary tree. The left child has larger probability than the right child.
2) The parent node has probability of the sum of the two symbols. The two child symbols are removed from the entries and replaced by their parent.
3) Recursively repeat step 1 until all symbols have been put in the binary coding tree. [2]

*Example:* Consider the word HELLO. Now we use this text in order to show how the algorithm works. Fig. 1 shows the probability of occurrence of a single alphabet. Now based on the algorithms, we go ahead with these alphabets probabilities. Fig. 2 represents the steps towards encoding the example word based of Huffman coding.

Table I shows the results after applying Huffman coding on the example. As shown in the table, the total number of bits is 10, the average bit number of each symbol is 10/5 = 2 bit/symbol.

*Effectiveness:* The Huffman algorithm is quite effective for compression in almost all file formats. Different versions of this algorithm can be designed for compression depending on the different types of applications. These versions mainly differ in the method of creating the Huffman tree.

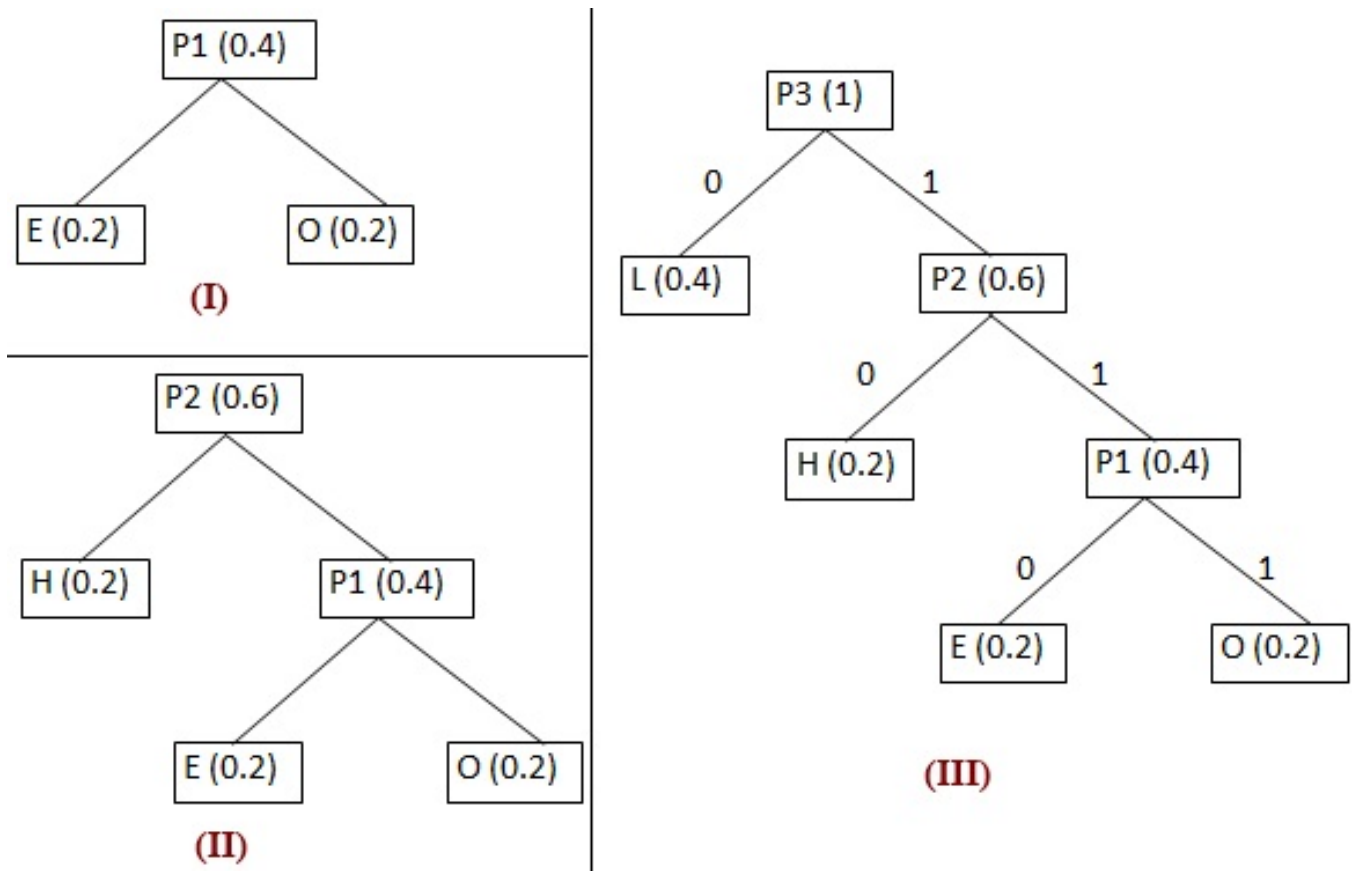

Fig. 2. Steps of applying huffman coding on HELLO

TABLE I
HUFFMAN ALGORITHM CODES FOR THE EXAMPLE

| Symbol | Count | Code | Number of bits |
|---|---|---|---|
| L | 2 | 0 | 1*2 |
| H | 1 | 10 | 2 |
| E | 1 | 110 | 3 |
| O | 1 | 1110 | 3 |

*Lemma:* The Huffman coding is a "minimum-redundancy code" algorithm and it has been proven that it is an optimal code for a set of data with given probability. It has the following properties: The two symbols with least probability are located at leaf of the same parent and their codes have same length. Symbol with higher probability has shorter code (fewer bits) than symbols with lower probability. The average length of code agrees with the following equation: [3]

$$n \leq averagelength < n+1 : n = \sum P_i log_2(1/P_i)$$

**Adaptive Huffman Coding**

In practical uses, we don't have the statistical information in prior. Adaptive Huffman method updates the statistical information while receiving new data. When the probability of a symbol increases (decreases), the code of that symbol would be updated in order to make it shorter (longer).

In the adaptive Huffman coding procedure, none of transmitter and receiver has information about the statistics of the source sequence at the start of transmission. The tree in the transmitter and the receiver consists of a single node which actually stands for all symbols not yet transmitted (NYT) and has a weight of zero. As we go ahead with transmission, nodes corresponding to symbols transmitted will be added to the tree, and the tree would be reconfigured using an update procedure. Before the beginning of transmission, a fixed code for each symbol is agreed between transmitter and receiver. [4] The algorithm goes this way: Select a conventional used code such as ASCII as an initial code. Whenever a symbol input has been occurred, the encoder checks the Huffman encoding tree. If there are any probabilities which are larger than their parents' generation, swap them, until all the probability of each entry is smaller than its parents' generation.

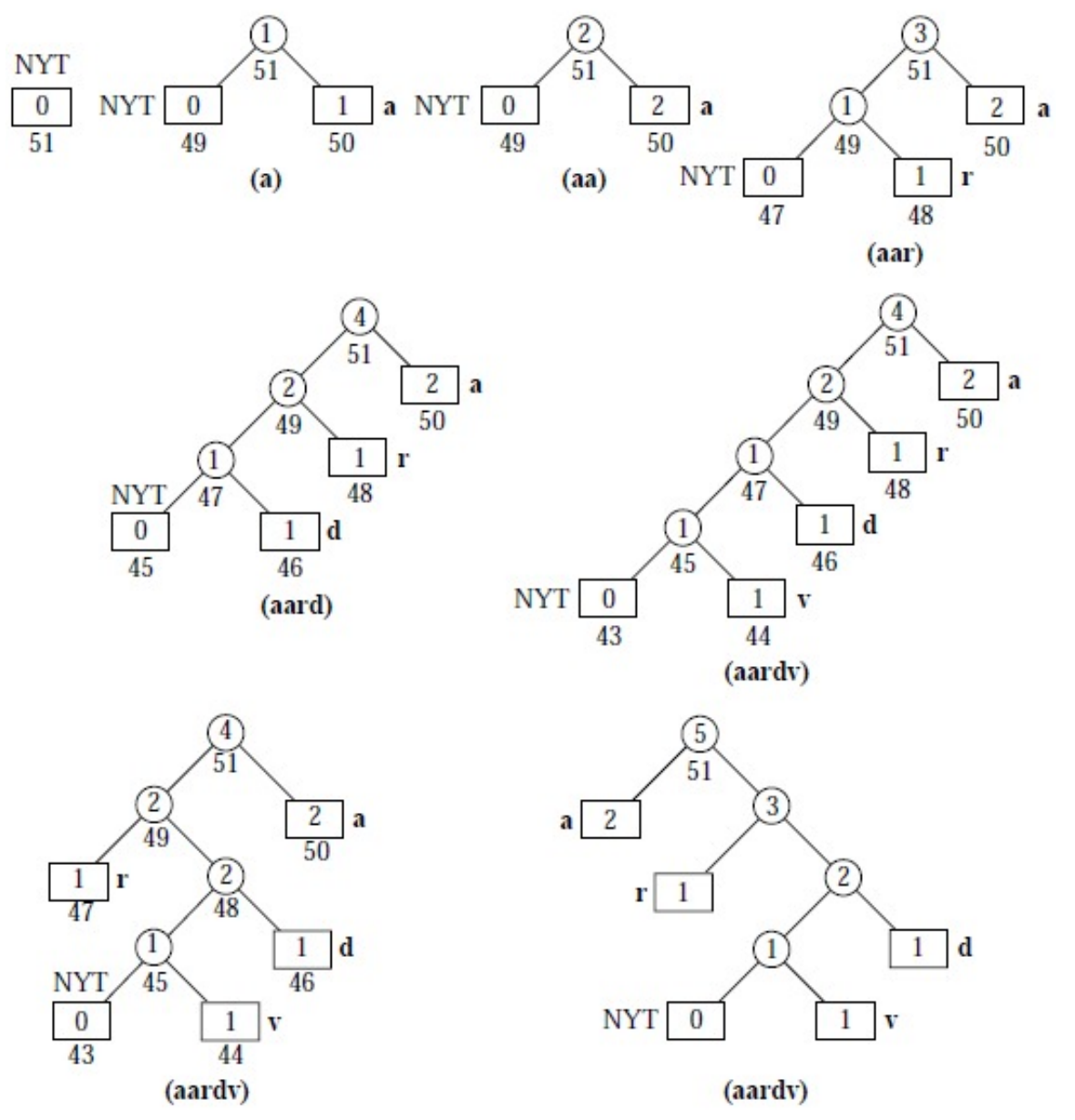

Fig. 3. Adaptive Huffman Coding after processing [a a r d v a r k]

As an example, assume we are encoding the message [a a r d v a r k], where our alphabet consists of the 26 lower-case letters of the English alphabet. Fig. 3 shows the coding process for our example.

Besides above methods, in some particular conditions, there are other effective methods. For instance Tunstall coding is a popular method which we would discuss in the followings.

**Tunstall Coding:**
Most of the variable-length coding algorithms that we look at in this survey encode letters from the source alphabet using codewords with varying numbers of bits: codewords with fewer bits for alphabets that occur more frequently and codewords with more bits for alphabets that occur less frequently. The Tunstall code is an important exception. The lengths of Tunstall Codes are equal, but the codes represent variable length of symbols. The main advantage of this method is that there is no accumulated error due to variable-length code. Instead of assigning fewer bits code to higher probability symbol, Tunstall coding concatenate symbol having highest probability with every other symbol together to form new strings replacing the original symbol. After K times iterations, the number of symbols becomes N+K(N-1). For an n-bit coding, $N + K(N - 1)2^n$. [**?**] Here we provide an example: We assume having a data set with the alphabets A, B, and C and we assume the probability distributions are: P(A)=0.6, P(B)=0.3, P(C)=0.1 . Alphabet A has the highest probability, so we find all two-alphabet strings with prefix A (AA, AB, and AC) to replace A. As figure 8 shows, after two runs, there are 7 entries $2^n$ . The final 3-bit code is shown in table II. [3]

TABLE II
RESULTS OF USING TUNSTALL CODING

| sequence | code |
|---|---|
| B | 000 |
| C | 001 |
| AB | 010 |
| AC | 011 |
| AAA | 100 |
| AAB | 101 |
| AAC | 110 |

**Applications of Huffman coding:**
Huffman coding and its different variations have gained so much importance due to wide variety of applications.

Arithmetic coding (which I will discuss later in this survey) can be viewed as a generalization of Huffman coding, in the sense that they produce the same output when every symbol has a probability of the form 1/2k; in particular this algorithm offers significantly better compression for small alphabet sizes. However Huffman coding algorithm remains in wide use because of its simplicity and high speed. Intuitively, arithmetic coding can offer better compression than Huffman coding because its ”code words” can have bit lengths which are not necessarily integers, while we see code words in Huffman coding can only have an integer number of bits. Therefore, there is an inefficiency in Huffman coding where a code word of length k only optimally matches a symbol of probability 1/2k and other probabilities are not represented as optimally; wihle the code word length in arithmetic coding can be made to exactly match the true probability of the symbol. [4]

Huffman coding today is often used as a "back-end" to some other compression methods. *DEFLATE* (PKZIP's algorithm) and multimedia codecs such as JPEG (which we would discuss later in this survey) and MP3 format have a model and quantization followed by Huffman coding. Image compression, Text compression and Audio compression are the major applications of Huffman coding:

- Image compression:

For instance, a simple application of Huffman coding to image compression would be to generate a Huffman code for the set of values that any pixel may take. For monochrome images, this set usually consists of integers from 0 to 255.

- Text compression:

Text compression seems natural for Huffman coding. In text, we have a discrete alphabet that has relatively stationary probabilities. For example, the probability model for a particular novel will not differ significantly from the probability model for another novel. Similarly, the probability model for a set of Java codes also is not going to be much different than the probability model for a different set of Java codes.

- Audio compression:

Another class of data that is very suitable for compression is CD-quality audio data. The audio signal for each stereo channel is sampled at 44.1 kHz, and each sample is represented by 16 bits. This means that the amount of data stored on a single CD is so much. If we want to transmit this data, the amount of channel capacity required would be significant. Compression woudl be definitely useful in this case.

### B. *Run-Length Coding*

Consider the idea say that once a pixel takes on a particular color (black or white), it is highly likely that the following pixels will also be of the same color. So, rather than code the color of each pixel separately, we can simply code the length of the runs of each color. Run-length encoding (RLE) is a very simple form of data compression in which sequences of data (which is called a run, repeating string of characters) are stored in two parts: a single data value and the count, rather than as the original run. This is most useful on data that contains many such runs: for example, simple graphic images such as icons, line drawings, and animations. It is not useful with files which don't have many runs as it could greatly increase the file size.

As an example, consider a plain black text on a solid white background. There will be many long runs of white pixels in the blank space, and many short runs of black pixels within the text. Let's consider a single line of pixels, with B representing a black pixel and W representing white: *WWWWWBWWWWWWBBB-WWWWWBWWWWW*

If we apply the run-length coding (RLE) data compression algorithm to the above pixel line, we get the following sequence: 5W1B5W3B5W1B5W This is to be interpreted as five Ws, one B, five Ws and so on.

The run-length code represents the original 25 characters in form of only 14 characters. Of course, the actual format used for the storage of images is generally binary rather than ASCII characters like this, but the principle remains the same. Even binary data files can be compressed with this method; file format specifications often cause repeated bytes in files. However, some newer compression methods use a generalization of RLE that can take advantage of runs of strings of characters instead of single characters (such as BWWBWWBWW).

And regarding its applications, since run-length encoding performs lossless data compression, it is well suited to palette-based images, like textures. It is not usually applied on realistic images such as photographs. Although JPEG (which will be discussed later in this survey) uses it quite effectively on the coefficients that remain after transforming and quantizing image blocks. [4]

Run-length encoding is used in fax machines. It is relatively efficient because most faxed documents have so much white space, with occasional interruptions of black.

### C. *Lempel-Ziv Algorithm*

The Lempel-Ziv algorithm which is a dictionary-based coding algorithm (unlike the previous coding algorithms which were based on probability coding) is the most preferred method for lossless file compression. This is mainly due to its adaptability to different file formats. It searches for frequently occurring phrases, say combinations of symbols and replaces them by a single symbol. It maintains a dictionary of these phrases and appends it to the final file. The length of the dictionary is fixed to a certain value. For small files, the length of the dictionary may exceed the length of the original file, but for large files, this method is very effective. [5]

The two main variants of the algorithm were described by Ziv and Lempel in two separate papers in 1977 and 1978, and are often refered to as LZ77 and LZ78. The algorithms differ in how far back they search and how they find matches. The LZ77 algorithm is based on the idea of a sliding window. The algorithm only looks for matches in a window within a fixed distance back from the current position. Gzip, ZIP, and V.42bis (a standard modem protocol) are all based on LZ77. The LZ78 algorithm is based on a more conservative approach to adding strings to the dictionary. [4]

Lempel-Ziv-Welch (LZW) is the mostly used Lempel-Ziv algorithm. Lempel-Ziv-Welch compression algorithm is proposed by Terry Welch in 1984. It is improved base on the LZ77 and L78 compression algorithm. The encoder builds an adaptive dictionary to represent the variable-length strings Without any prior probability information. The decoder also builds the same dictionary in encoder according to received code dynamically. In text data, some symbols/characters occur together frequently. The encoder can memorize these symbols and represent them in one code. [6] A typical LZW code is 12-bits length (4096 codes). The first 256 (0 255) entries are ASCII codes, to represent individual character. The other 3840 (256 4095) codes are defined by encoder to represent variable-length strings. Now LZW is applied in GIF images, UNIX compress, and others. [2] The algorithm goes this way:

1) Initialize the dictionary.
2) Combine symbols of the input data together sequentially into buffer until the longest string/symbol can be found in the dictionary.
3) Send the code representing in the buffer.
4) Save the string in the buffer combining with the next symbol in the next empty code into the dictionary.
5) Clear the buffer, then repeat Steps 2 to 5 until the end of total data.

An example of LZW is shown in table III. [4] The input string is ABABBABCABABBA and the initial code is 1, 2, and 3 to representing A, B, and C.

The encoded string is ”124523461”. The data is compressed from 14 characters to 9 characters. The compression ratio is thus 14/9 = 1.56. If we also take into account the bit number, a typical 12-bits LZW for example, the ratio becomes (814)/( 129)=1.04.

**Applications of LZ coding:**

TABLE III
RESULTS OF USING TUNSTALL CODING

| Symbol | Next symbol | Output | Code | String |
|---|---|---|---|---|
| A | B | 1 | 4 | AB |
| B | A | 2 | 5 | BA |
| AB | B | 4 | 6 | ABB |
| B | A | | | |
| BA | B | 5 | 7 | BAB |
| B | C | 2 | 8 | BC |
| C | A | 3 | 9 | CA |
| A | B | | | |
| AB | A | 4 | 10 | ABA |
| A | B | | | |
| AB | B | | | |
| ABB | A | 6 | 11 | ABBA |
| A | EOF | 1 | | |

Since the publication of Terry Welchs article, there has been a steadily increasing number of applications that use some variant of the LZ algorithm. Among the LZ variants, the most popular is the LZW algorithm. However LZW algorithm was initially the mostly used algorithm, patent concerns has led to increasing use of the LZ algorithm. The most popular implementation of the LZ algorithm is the *deflate* algorithm initially designed by Phil Katz. *Deflate* is a lossless data compression algorithm that uses a combination of the LZ algorithm and Huffman coding. So as we discussed in the Huffman coding section, it is considered a variant of both algorithms. It is part of the popular zlib library developed by Jean-loup Gailly and Mark Adler. [7] Jean-loup Gailly also used deflate in the widely used gzip algorithm. The deflate algorithm is also used in PNG image format which we will also describe below. [4] Here we enumerate some of major applications of LZ algorithms:

- File CompressionUNIX compress:

The UNIX compress command is one of the earliest applications of LZW. The size of the dictionary is adaptive. We start with a dictionary of size 512. This means that the transmitted codewords have 9 bits in length. Once the dictionary has filled up, the size of the dictionary is doubled to 1024 entries. The codewords transmitted at this point have 10 bits. The size of the dictionary is progressively doubled as it fills up. In this way, during the earlier part of the coding process when the strings in the dictionary are not very long, the codewords used to encode them also have fewer bits. The maximum size of the codeword, bmax, can be set by the user to between 9 and 16, with 16 bits being the default. Once the dictionary contains 2bmax entries, compress becomes a static dictionary coding technique. At this point the algorithm monitors the compression ratio. If the compression ratio falls below a threshold, the dictionary would be flushed, and the dictionary building process is restarted. This way, the dictionary always reflects the local characteristics of the source.

- Image Compression-GIF format:

The Graphics Interchange Format (GIF) was developed by Compuserve Information Service to encode graphical images. It is another implementation of the LZW algorithm and is very similar to the compress command in Unix, as we mentioned in the previous application.

- Image CompressionPNG format:

PNG standard is one of the first standards to be developed over the Internet. The motivation for its development was an announcement in Dec. 1994 by Unisys (which had obtained the patent for LZW from Sperry) and CompuServe that they would start charging royalties to authors of software which included support for GIF. The announcement resulted in a say revolution in data compression that formed the core of the Usenet group comp.compression. The community decided that a patent-free replacement for GIF should be developed, and within three months PNG was born. (For more detailed history of PNG as well as software and much more, go to the PNG website maintained by Greg Roelof, http://www.libpng.org/pub/png/.)

- Compression over ModemsV.42

The ITU-T Recommendation V.42 bis is a compression standard developed for using over a telephone network along with error-correcting procedures described in CCITT Recommendation V.42. This algorithm is used in modems connecting computers to remote users. The algorithm operates in two modes, a transparent mode and a compressed mode. In the transparent mode, the data are transmitted in uncompressed form, while in the compressed mode a LZW algorithm is used for compression. [4] [2] [3]

### D. *Arithmetic Coding*

Arithmetic Coding is another coding algorithm based on entropy encoding. Unlike Huffman coding which uses variable-length codes to represent each symbols, Arithmetic Coding encode the whole data into one single number within the interval of 0 and 1. The algorithm implement by separating 0 to 1 into segments according to the number of different symbols. The length of each segment is proportional to the probability of each symbol. Than the output data is located in the corresponding segment according to the symbol. [8]

TABLE IV
USING ARITHMETIC CODING

| symbol | probability | interval |
|---|---|---|
| A | 0.2 | [0, 0.2) |
| B | 0.1 | [0.2, 0.3) |
| C | 0.2 | [0.3, 0.5) |
| D | 0.05 | [0.5, 0.55) |
| E | 0.3 | [0.55, 0.85) |
| F | 0.05 | [0.85, 0.9) |
| @ | 0.1 | [0.9, 1) |

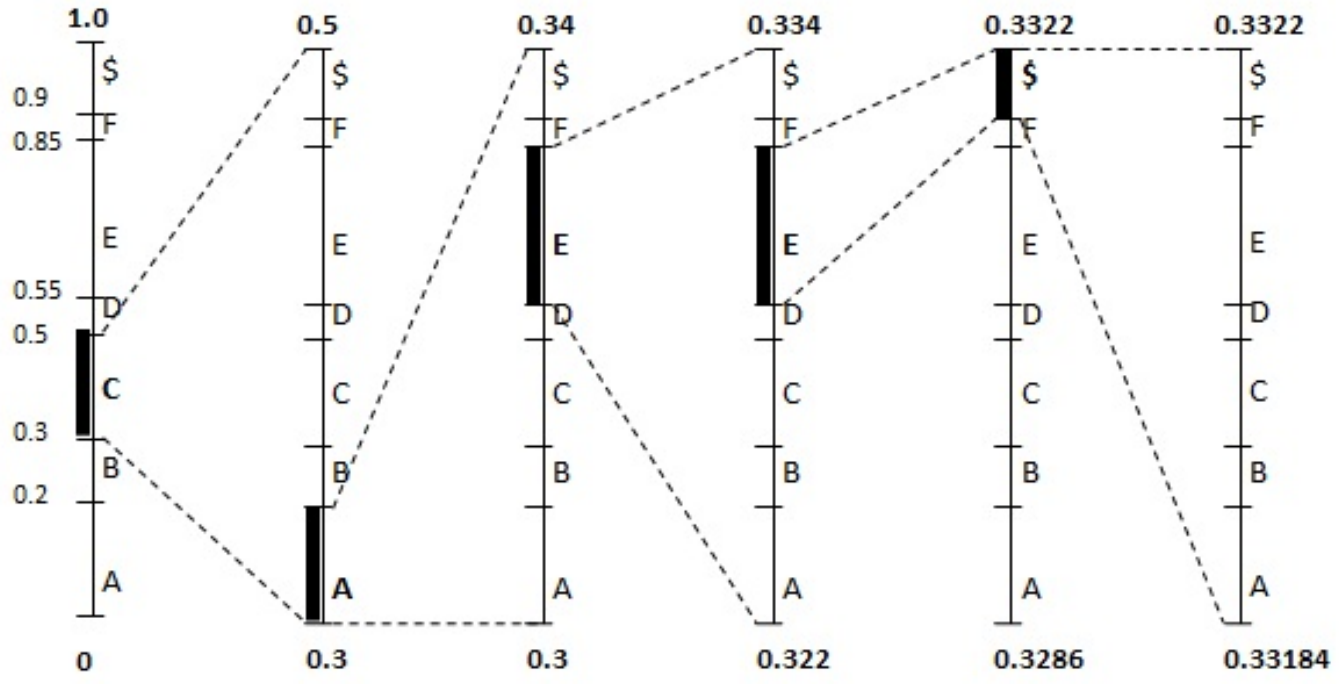

Fig. 4. Arithmetic coding process for our example

Consider an alphabet set A, B, C, D, E, F, @, where @ represent the end of message. The probability and segment range is shown in table IV.

Now we take a string CAEE@ for example. The initial interval [0, 1) is divided into seven interval, corresponding to the number of symbols. The first symbol is C, which locate in the third segment, [0.3, 0.5). Then continuously partition interval [0.3, 0.5) into seven segments. The second symbol A is located in the first segment interval from 0.3 to 0.34. Continuously repeat partition until the end of string. The encoding step is illustrated in Fig. 4. [4]

The final interval is bounded by 0.33184 and 0.33320. The range is also the probability of the string. Interval = $P_C P_A P_E P_E P_@$=0.2*0.2*0.3*0.3*0.1 = 0.00036. If we calculate $log_2 1/0.00036$ it would be almost 11.4. Thus using 12 bits, we can encode this number.

The final step is to pick an arbitrary number in the range [0.33184, 0.33320), such as 0.332, or 0.33185. When binary coding, a shortest binary fractional number is the best suggested. Therefore the code 0.01010101 is generated here, which would be equal to 0.3320.

**Applications:** Arithmetic coding is used in a variety of lossless and lossy compression applications. It is a part of many international standards. In the area of multimedia there are some organizations that develop standards. The International Standards Organization (ISO) and the International Electrotechnical Commission (IEC) are industry groups that work on multimedia standards, while the International Telecommunications Union (ITU) works on multimedia standards. Quite often these institutions work together to create international standards. In the following of our survey, we will see how arithmetic coding is used in JPEG, as an example of image compression.

## IV. Lossy Data Compression

Lossy compression is compression in which some of the information from the original message sequence is lost. This means the original sequences cannot be regenerated from the compressed sequence. Just because information is lost doesn't mean the quality of the output is reduced. For example, random noise has very high information content, but when present in an image or a sound file, we would typically be perfectly happy to drop it. Also certain losses in images or sound might be completely undetectable to a human viewer (e.g. the loss of very high frequencies). For this reason, lossy compression algorithms on images can often get a better compression ratio by a factor of 2 than lossless algorithms with an undetectable loss in quality. However, when quality does start degrading in an observable way, it is important to make sure it degrades in a way that is least objectionable to the viewer (e.g., dropping random pixels is probably more unpleasant than dropping some color information). For these reasons, the ways most lossy compression techniques are used are highly dependent on the media that is being compressed. Lossy compression for sound, for example, is very different than lossy compression for images.

In this section, I will go over two significant techniques which can be applied to various contexts, and in the next sections we will see some of their major applications in the real world.

### A. *Vector and Scalar Quantization*

Quantization is the procedure of constraining something from a relatively large or continuous set of values (such as the real numbers) to a relatively small discrete set (such as integers).

Scalar quantization is one of the simplest and most general idea for lossy compression. Scalar quantization is a mapping of an input value x into a finite number of output values, y. Many of the fundamental ideas of quantization and compression are easily introduced in the simple context of scalar quantization. Fig. 5 shows examples of uniform and non uniform quantizations. [2]

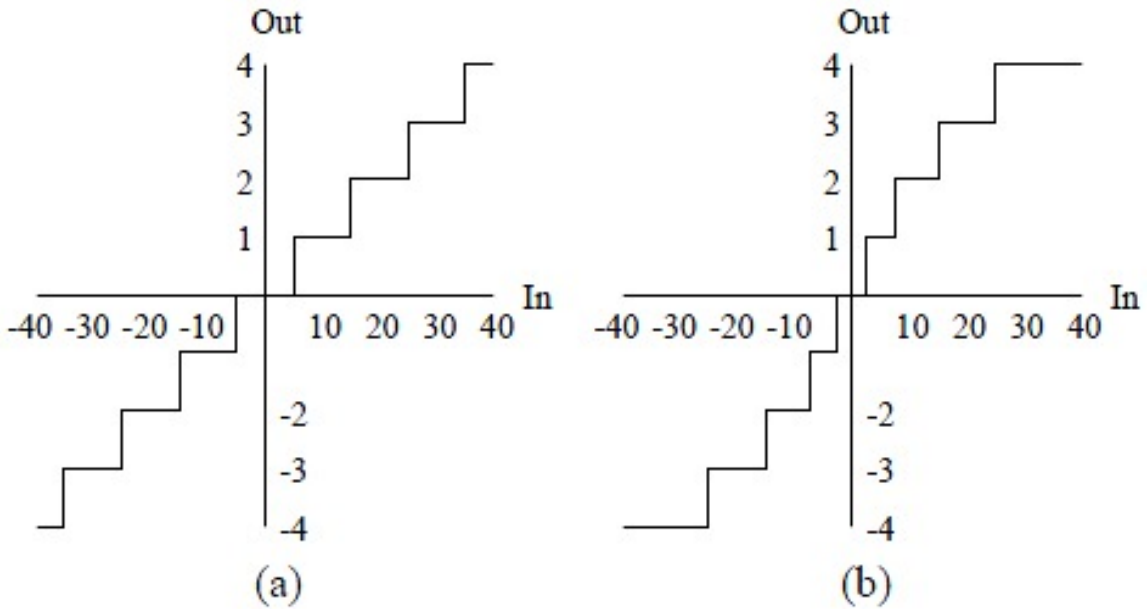

Fig. 5. Examples of uniform (a) and non-uniform (b) scalar quantization.

Vector quantization is a different type of quantization, which is typically implemented by selecting a set of representatives from the input space, and then mapping all other points in the space to the closest representative. The representatives could be fixed for all time and part of the compression protocol, or they could be determined for each file (message sequence) and sent as part of the sequence. The most interesting aspect of vector quantization is how one selects the representatives. Typically it is implemented using a clustering algorithm that finds some number of clusters of points in the data. A representative is then chosen for each cluster by either selecting one of the points in the cluster or using some form of centroid for the cluster. Finding good clusters is a whole interesting topic on its own. [9]

Vector quantization is most effective when the variables along the dimensions of the space are correlated. Fig. 6 shows an example of possible representatives for a height-weight chart. There is clearly a strong correlation between people's height and weight and therefore the representatives can be concentrated in areas of the space that make physical sense, with higher densities in more common regions. Using such representatives is very much more effective than separately using scalar quantization on the height and weight. [4] Vector quantization, as well as scalar quantization, can be used as part of a lossless compression technique. In particular if in addition to sending the closest representative, the coder sends the distance from the point to the representative, then the original point can be reconstructed. The distance is often referred to as the residual. In general this action alone would not cause compression, but in case the points have been tightly clustered around the representatives,

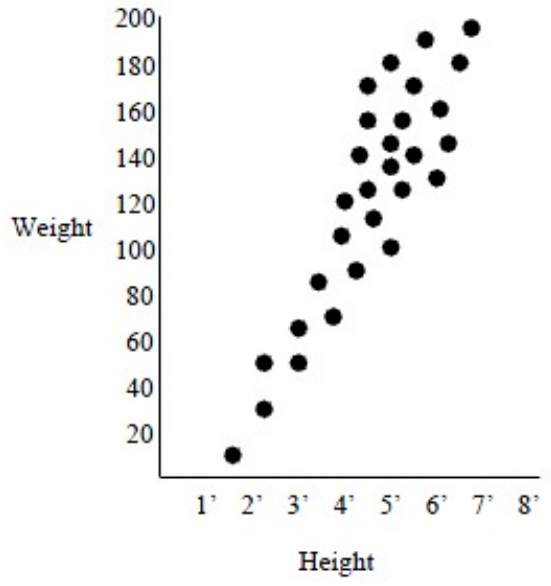

Fig. 6. Example of vector quantization.

then this approach can be very effective for lossless data compression.

## V. A Case Study: Lempel-Ziv vs Huffman Coding

Usually data compression algorithms do not take into account the type of data they are compressing. They can be applied to computer data files, documents, images, etc. In this section I will discusses two of the most wide used algorithms for data compression as Ive mentioned before: Huffman coding and Lempel-Ziv coding (Ive run LZ 77 v01 coding algorithm).

In this section, I will run my simulation for these two coding algorithms, and try to have a comparison between them:

**Test 1:** ”Miss Issipi is living in Mississippi . but she misses Mississippi river everyday .”
**Test 2:** ”Hello. I’m Mohammad. I’m running the simulation for two coding algorithms.”

### *Simulation Analysis*

Fig. 7 shows our simulation results for two test data. We used Java programming language for simulation.

As we know, Huffman coding uses a variable length code for each of the elements within the information. This normally involves analyzing the information to determine the probability of elements within the information. The most probable elements are coded with a few bits and the least probable coded with a greater number of bits.

The great advantage of Huffman coding is that, although each character is coded with a different number of bits, the receiver will automatically determine the character whatever their order.

But Lempel-Ziv algorithm builds a dictionary of frequently used groups of characters (or 8-bit binary values). Before the file is decoded, the compression dictionary must be sent (if transmitting data) or stored (if data is being stored). This method is good at compressing text files because text files contain ASCII characters (which are stored as 8-bit binary values) but not so good for graphics files, which may have repeating patterns of binary digits that might not be multiples of 8 bits.

| | | **Lempel-Ziv** | **Huffman** |
|---|---|---|---|
| | Raw data length (in bits) | 664 | 664 |
| **Test 1** | encoded data length (in bits) | 576 | 304 |
| | Compression factor | 1.152777778 | 2.184210526 |
| | Raw data length (in bits) | 592 | 592 |
| **Test 2** | encoded data length (in bits) | 640 | 320 |
| | Compression factor | 0.925 | 1.85 |

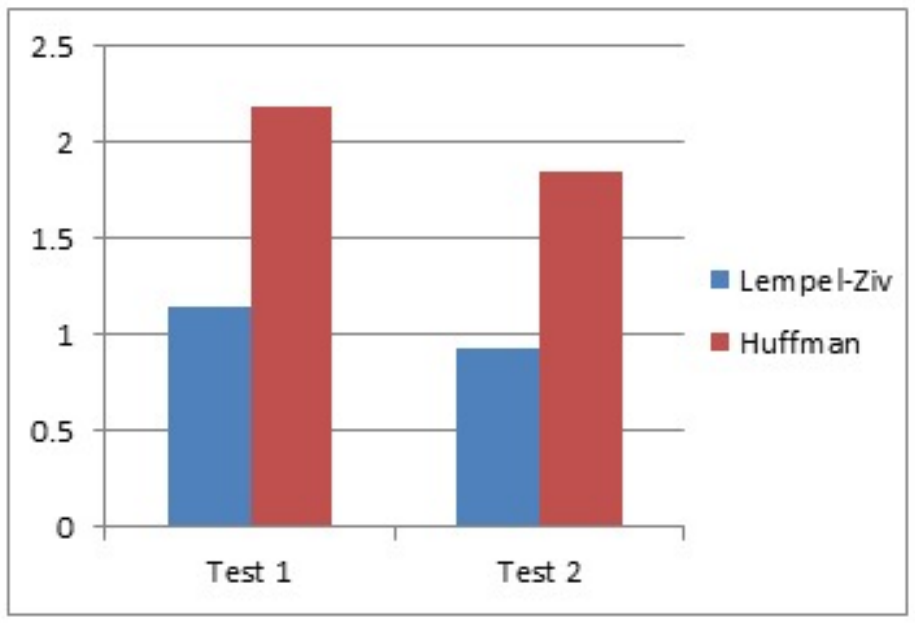

Fig. 7. Simulation results of comparison between Huffman and LZ algorithms based on our two tests.

The simulation above shows lesser compression for LZ than the Huffman Algorithm, but for large files, the LZ algorithm is found to be more efficient. It can also adapt to different file formats better. LZ algorithm also has a lot of variations. Two distinct forms of the LZ algorithm are LZ77 and LZ78. They have also evolved into further forms. Popular forms are LZW, LZH, LZC etc. The LZ algorithm is the more adaptive of the two to different file formats, although both the algorithms can be suitably varied to achieve better compression.

The LZ algorithm can also be used in continuous data transmission, where the receiver maintains pointers to the previously encountered patterns and can thus decode constantly. This is a very useful feature of the LZ algorithm.

The Huffman algorithm is better suited to smaller files, although compression is not significant for very small files.

On the whole, the LZ algorithm is preferred over the Huffman algorithm, and a number of popular file compression utilities use variations of this algorithm.

Please note that it is rather obvious that the more the words in a test data are similar, the more the compression ratio is. This fact is also observable from our simulation results.

## VI. Multimedia Compression: JPEG and MPEG

Multimedia images have become a vital and ubiquitous component of everyday life. The amount of information encoded in an image is quite large. Even with the advances in bandwidth and storage capabilities, if images were not compressed many applications would be too costly.

The JPEG and the related MPEG format are examples of multimedia compression. They are used very widely in practice, and also they use many of the compression techniques we have been talking about throughout this survey, including Huffman codes, arithmetic codes, run length coding and scalar quantization.

JPEG is used for still images and is the standard used on the web for photographic images (the GIF format is often used for textual images). MPEG is used for video, based on a variant of JPEG (i.e. each frame is coded using a JPEG variant). Both JPEG and MPEG are lossy formats.

### A. *Image Compression: JPEG Compression Algorithm*

Image compression is the process of reducing the amount of data required to represent a digital image. This is done by removing all redundant or unnecessary information. An uncompressed image requires an enormous amount of data to represent it.

JPEG is the image compression standard developed by the Joint Photographic Experts Group. JPEG is a lossy compression algorithm that has been conceived to reduce the file size of natural, photographic-like true-color images as much as possible without affecting the quality of the image as experienced by the human JPEG works on either full-color or gray-scale images; it does not handle bi-level (black and white) images, at least not well. It doesn't handle color mapped images either; you have to pre-expand those into an unmapped full-color representation. JPEG works best on ”continuous tone” images. Images with many sudden jumps in color values will not compress well. There are a lot of parameters to the JPEG compression process. By adjusting the parameters, you can trade off compressed image size against reconstructed image quality. [10]
In the followings we will discuss the steps used for JPEG coding algorithm:

**1. Color space conversion**

the best compression results are achieved if the color components are independent (noncorrelated), such as in YCbCr, where most of the information is concentrated in the luminance and less in the chrominance. RGB color components can be converted via a linear transformation into YCbCr components as the equation below shows: [10]

$$\begin{bmatrix} Y \\ Cb \\ Cr \end{bmatrix} = \begin{bmatrix} 0.299 & 0.587 & 0.144 \\ -0.159 & -0.332 & 0.050 \\ 0.500 & -0.419 & -0.081 \end{bmatrix} x \begin{bmatrix} R \\ G \\ B \end{bmatrix}$$

**2. Chrominance downsampling**

Using YCbCr color space, we can also save space by compressing resolution of Cb and Cr components. They are chrominance component and we can reduce them in order to make the image compressed. Because of the importance of luminance for the eyes, we do not need the chrominance to be as frequent as luminance, so we can downsample it. Thus we can remove some of Cb and Cr elements. As a result, for instance transforming RGB, 4:4:4 format into YCbCr 4:2:2, we would gain a data compression of ratio 1.5. However this step is considered as an optional process.

**3. Discrete cosine transform (DCT):**

At this step, each component (Y, Cb, Cr) of every 88 block is then converted to a frequency domain representation. The DCT equation is a complicated equation, having two cosine coefficients. We do not mention here, for more details refer to JPEG standard.

**4. Quantization:**

As we mentioned in the lossy compression section, for human eye, luminance is more important than chrominance. For eyes, seeing small differences in brightness within a large area is more distinguishable than the exact strength of a high frequency brightness variation. Using this property, we can greatly reduce information in the high frequency components. JPEG coding does that by simply dividing every components in the frequency domain by a constant for that component, and then rounding to the nearest integer. So as a result, many of the higher frequency components are rounded to zero, and most of the rest components become small positive or negative numbers, taking fewer bits for storage. The main lossy method in the whole process is done at this step. The standard quantization for JPEG is provided in table V. [**?**] However many other tables have been also defined and used.

**5. Entropy coding:**

Entropy coding is a type of lossless data compression. Here we order image components in a *zigzag* form, then using run-length encoding (RLE) algorithm that joins similar frequencies together to compress the sequence. We discusses RLE coding in the lossless compression section.

TABLE V
STANDARD QUANTIZATION USED BY JPEG

| | | | | | | | |
|---|---|---|---|---|---|---|---|
| 16 | 11 | 10 | 16 | 24 | 40 | 51 | 61 |
| 12 | 12 | 14 | 19 | 26 | 58 | 60 | 55 |
| 14 | 13 | 16 | 24 | 40 | 57 | 69 | 56 |
| 14 | 17 | 22 | 29 | 51 | 87 | 80 | 62 |
| 18 | 22 | 37 | 56 | 68 | 109 | 103 | 77 |
| 24 | 35 | 55 | 64 | 81 | 104 | 113 | 92 |
| 49 | 64 | 78 | 87 | 103 | 121 | 120 | 101 |
| 72 | 92 | 95 | 98 | 112 | 100 | 103 | 99 |

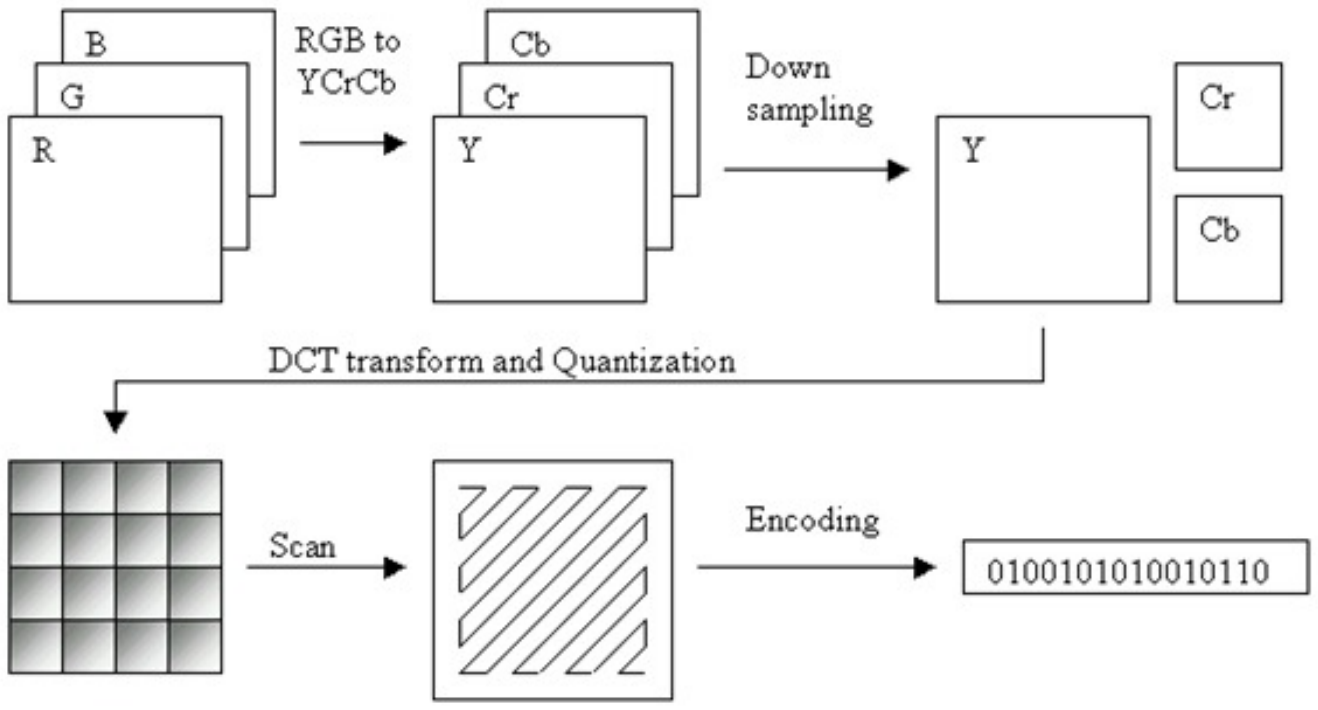

Fig. 8. JPEG compression steps

### 6. Huffman algorithm

After applying the previous steps, then the resulting data would be sequence of DCT coefficients. Now at this step which is the last step, we compress these coefficients by using a Huffman coding or a arithmetic compression algorithm. Mostly Huffman would be used which would be considered as the second lossless compression in this scheme.

In general, the algorithm compresses images in different steps. Fig. 8 shows these different steps used for compression. Many other compression approaches for image and video compression are similar to JPEG. Aslo many of the variants of JPEG use the basic concepts of JPEG scheme which are usually designed for some specific applications. [11] [12] [2]

## B. *Video Compression: MPEG Compression Algorithm*

Mostly people have some basic information about what MPEG compression is. It is used to compress video files. MPEG is the short form of Moving Pictures Expert Group, almost the same as JPEG. It is said that the founding fathers of MPEG are Leonardo Chairiglione (from Italy) and Hiroshi Yasuda (from Japan). The basic idea for MPEG compression algorithm is to transform a stream of discrete samples into a bit stream of tokens, taking less space. In theory, a video stream is a sequence of discrete images. MPEG uses the special or temporal relation between these successive frames for compression of video streams. This approach is the basic idea in many of the approaches we have seen in this survey. The more effectively a technique is in exploiting these relations in a piece of data, the more effective it would be for data compression.

Fig. 9. The overal idea of MPEG coding

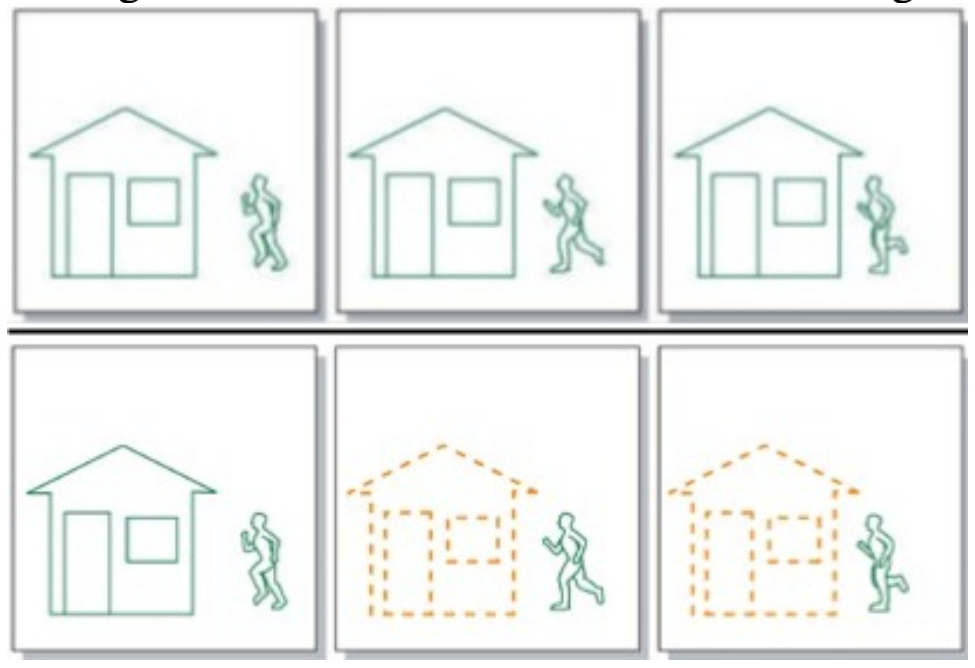

The first public standard of the MPEG committee was the MPEG-1, ISO/IEC 11172 that was first released in 1993. MPEG-1 video compression is based on the same approach used in JPEG. Besides that, it also leverage some methods for efficient coding of a video sequence. [13]

In MPEG coding algorithm, we only code the new parts of the video sequence along with the information of the moving parts of the video. As an example, consider having Figure 9, the upper three pictures. For compressing, we would just take into account the new parts, which is shown in Figure 9, the bottom three sequences. [4]

The basic principle for video compression is the image-to-image prediction. The first image in a sequence of images is I-frame. These frames shows beginning a new scene thus do not need to be compress since they have no dependency outside of that image. But the other frames may use part of the first image as a reference. An image that is predicted from one reference image is called a P-frame and an image which is bidirectionally predicted from two other reference images is called a B-frame. So overall, we would have the followings frames for MPEG coding:

*I-frames*: self-contained; no need to be compress
*P-frames*: Predicted from last I or P reference frame
*B-frames*: Bidirectional; predicted from two references one in the past and one in the future, the best matching would be considered.

Figure 10 shows a sequence of these frames.

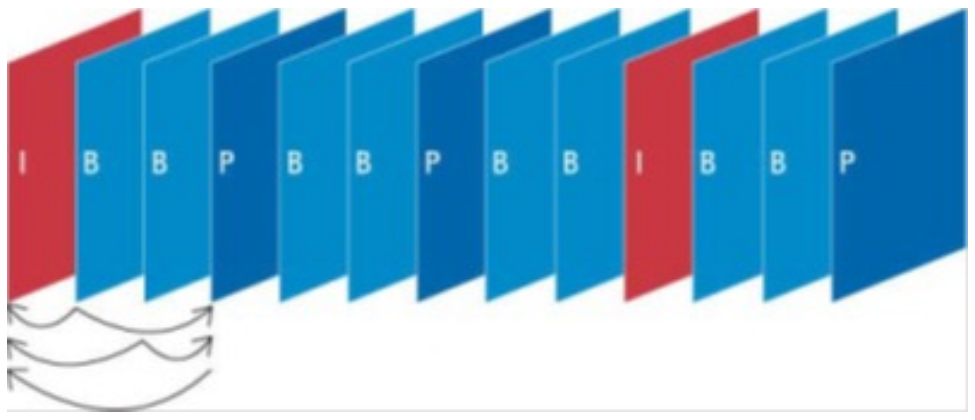

Fig. 10. A sequence of MPEG defined frames

Since I-frames are independent images, we compress them like a single image. The particular technique used by MPEG is a variant of the JPEG technique (the color transformation and quantization steps are somehow different). To decode any frame we only need to search and find the closest previous I-frame to the current frame and go from there. This is important for allowing reverse playback, skip-ahead, or error-recovery. [4]

As we mentioned, our purpose for coding P-frames is to find matching images, actually a set of pixels with similar patterns in the previous reference frame and then we would code just the difference between the P-frame and the match we have found. In order to find these matches, MPEG algorithm divides P-frame into 16x16 blocks.

Figure 11 shows the different processes used for encoding these different 16x16 blocks.. For each target block in the P-frame, the encoder finds a reference block in the previous P-frame or I-frame which is the best match for it. The reference block need not be aligned on a 16-pixel boundary and can potentially be anywhere in the image. Motion vector is the vector showing this direction. Once the match is found, the pixels of the reference block are subtracted from the corresponding pixels in the target block. This gives a residual which is very close to zero. This residual is coded using a similar approach to JPEG algorithm. In addition to sending the coded residual, the coder also needs to send the motion vector, which is coded using Huffman compression algorithm.

The motivation for searching other locations in the reference image for a match is to allow for the efficient encoding of motion. In particular if there is a moving object in the sequence of images (consider a playing avatar), or if the camera is just moving, then the best match will not be in the same location in the image. It should be noted that if no good match is found, then the block would be coded the same as an I-frame. Practically, this searching procedure for finding a good match for each block is the most computationally costly part of MPEG encoding. With current technology, real-time MPEG encoding is only possible with the help of powerful hardware devices. However, the decoder part is cheap since the decoder is given the motion vector and only needs to look up the block from the previous image.

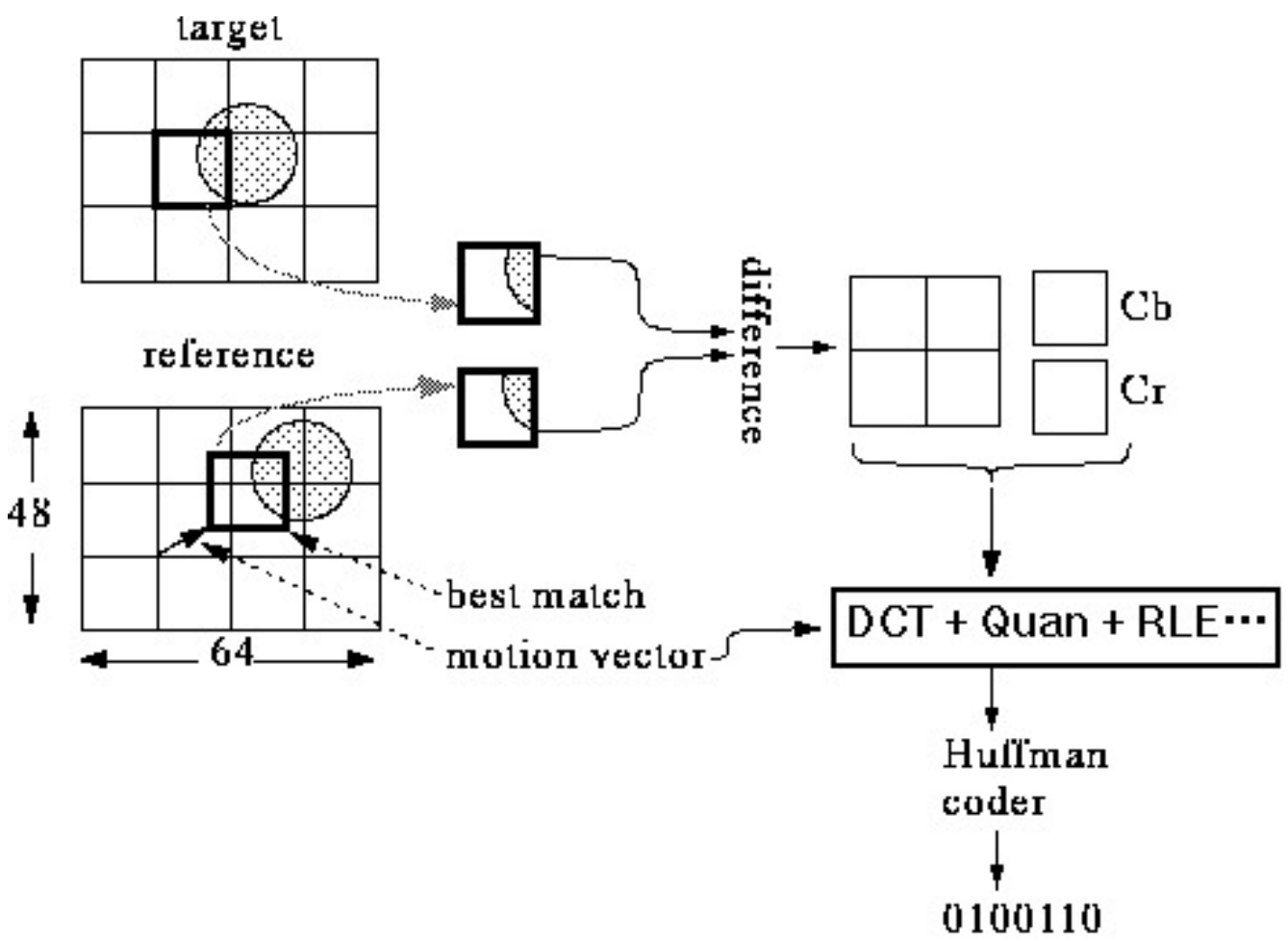

Fig. 11. MPEG overal coding process

Actually for B-frames, we look for reusable data in both directions. The general approach is similar to what we use in P-frames, but instead of just searching in the previous I- or P-frame for a match, we would also search the next I- or P-frame. If a good match is found in each of them, then we would take an average of the two reference frames. If only one good match is found, then this one would be used as the reference. In these cases, the coder needs to send some information saying which reference has been used. [2] [2]

**MPEG compression evaluation:**
Here we evaluate the effectiveness of MPEG coding algorithm using a real world example. We can examine typical compression ratios for each frame type, and form an average weighted by the ratios in which the frames are typically interleaved.

Starting with a 356260 pixel, 24-bit color image, typical compression ratios for MPEG-I are provided here:

| Type | Size | Ratio |
|---|---|---|
| I | 18 Kb | 7:1 |
| P | 6 Kb | 20:1 |
| B | 2.5 Kb | 50:1 |
| Avg | 4.8 Kb | 27:1 |

If one 356 260 frame requires 4.8 Kb, then for providing a good video at a rate of 30 frames/second, then MPEG requires: 30 frames/sec 4.8 Kb/frame 8 b/bit = 1.2 Mbits per second.

Thus far, we have been concentrating on the visual component of MPEG. Adding a stereo audio stream will require roughly another 0.25 Mbits/sec, for a grand total bandwidth of 1.45 Mbits/sec. This would be fit fine in the 1.5 Mbit per second capacity of a T1 line. In fact, this specific limit was a design goal in the MPEG design that time. Real-life MPEG encoders are adaptive; they track bit rate as they encode, and will dynamically adjust compression qualities to keep the bit rate within some user-selected bound. This bit-rate control can also be important in other contexts. For example, video on a multimedia CD-ROM must fit within the relatively poor bandwidth of a typical CD-ROM drive.

**MPEG Applications:**

MPEG has so many applications in the real world. We enumerate some of them here:

1. Cable Television. Some TV systems send MPEG-II programming over cable television lines.
2. Direct Broadcast Satellite. MPEG video streams are received by a dish/decoder, which extracts the data for a standard NTSC television signal.
3. Media Vaults. Silicon Graphics, Storage Tech, and other vendors are producing on-demand video systems, with twenty file thousand MPEG-encoded films on a single installation.
4. Real-Time Encoding. This is still the exclusive province of professionals. Incorporating special-purpose parallel hardware, real-time encoders can cost twenty to fifty thousand dollars. [4] [2] [3]

## VII. Today's Data Compression: Applications and Issues

A key in acceptance of data compression algorithms is finding an acceptable tradeoff between performance and complexity. For performance we have two factors that run counter to each other and require another compromise. These factors are the end-user perception of compression (e.g. image quality in image compression) and the data rate compression achieved. System complexity eventually defines the cost of encoding and decoding devices. Here we briefly discuss some todays data compression issues, and at the end will briefly discuss some research works towards energy efficiency which nowadays is the most concerning field of study as we are turning in to a green computing.

### A. *Networking*

Today, with the increasing number of users and teleworking along with emerging application deployment models which use cloud computing, causes additional pressure on existing network connections in because of more data are being transmitted.

One of the important and major roles of data compression is using them in computer networks. However achieving a high compression ratio is necessary for improving applications' performance on networks with limited bandwidth, system throughput also plays an important role. If the compression ratio is too low, the network will remain saturated and performance gains will be minimal. Similarly, if compression speed is too low, the compressor will become the bottleneck. Employee productivity can be dramatically impacted by slow networks that result in poorly performing applications.

Organizations have turned to network optimization as a way to combat the challenges of assuring application performance and help ensure timely transfer of large data sets across constrained network links. Many network data transfer optimization solutions are focused just on network-layer optimizations. Not only are these solutions are inflexible, they also fail to include optimizations that can further enhance the performance of applications transferring data over network links. [14]

### B. *Packet-based or Session-based Compression*

Many of network compression systems are packet-based. Packet-based compression systems buffer packets destined for a remote network with a decompressor. These packets are then compressed either within a single time or as a group and then sent to the decompressor where the process is reversed. Packet-based compression has been available for many years and can be found in routers, VPN clients.

Packet-based compression systems have additional problems. When compressing packets, these systems must choose between writing small packets to the network and performing additional work to aggregate and encapsulate multiple packets. Neither option produces optimal results. Writing small packets to the network increases TCP/IP header overhead, while aggregating and encapsulating packets adds encapsulation headers to the stream.

### C. *Dictionary Size for Compression*

One limitation of that almost all compression utilities have in common is limited storage space.

Some utilities, such as GNUzip (gzip), store as little as 64 kilobytes (KBs) of data. Others techniques, such as disk-based compression systems, can store as much as 1 TByte of data.

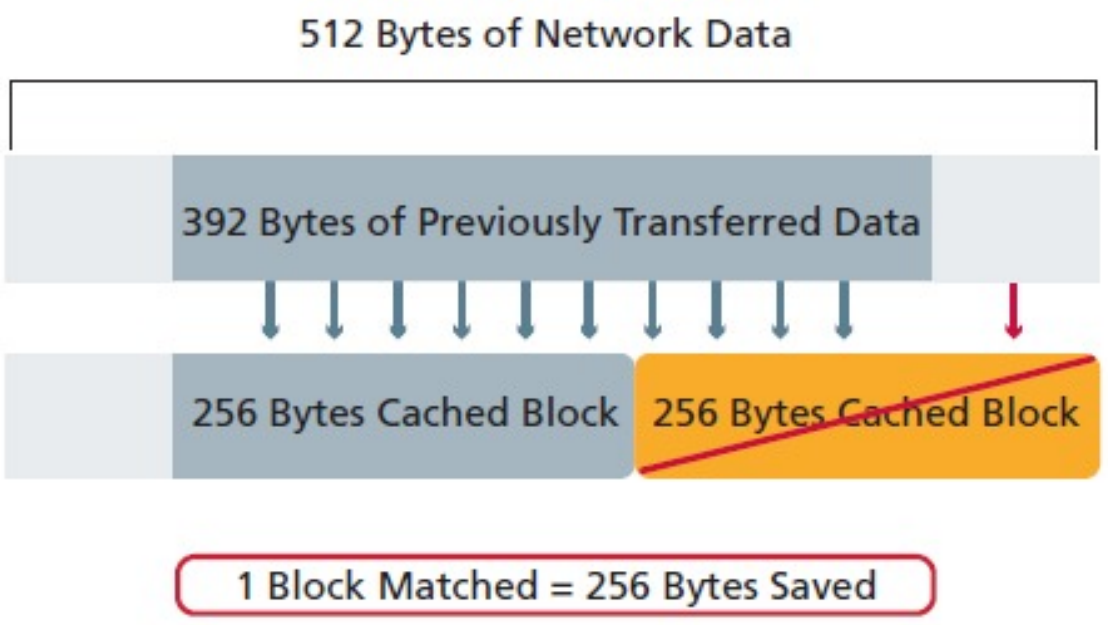

Fig. 12. Compressing 512B of data in a 256B-block system

Similar to requests to a website, not all bytes transferred on the network repeat with the same frequency. Some byte patterns occur with great frequency because they are part of a popular document or common network protocol. Other byte patterns occur only once and are never repeated again. The relationship between frequently repeating byte sequences and less frequently repeating ones is seen in Zipfs law. (%80 of requests are for %20 of data; data with higher ranks.)

### D. *Block-based or Byte-based Compression*

Block-based compression systems store segments of previously transferred data flowing across the network. When these blocks are encountered a second time, references to the blocks are transmitted to the remote appliance, which then reconstructs the original data.

A critical shortcoming of block-based systems is that repetitive data almost never is exactly the length of a block. As a result, matches are usually only partial matches, which do not compress some repetitive data. Figure 12 illustrates what happens when a system using a 256-byte block size attempts to compress 512 bytes of data. [14] [4]

### E. *Energy Efficiency*

The most recent works are being done in the area of energy efficiency, especially regarding Wireless Sensor Networks (WSNs). Just as a reminder that a wireless sensor network consists of spatially distributed sensors to monitor physical or environmental conditions, such as temperature, sound, vibration, pressure, motion or pollutants and to cooperatively pass their data through the network to a main location. Nowadays lots of papers and research works have been devoted to WSNs. Size and cost constraints on sensors result in constraints on resources on them such as energy, memory, computational speed and communications bandwidth. The vast sharing of data among the sensors needs energy efficiency, low latency and also high accuracy.

Currently, if sensor system designers want to compress acquired data, they must either develop application-specific compression algorithms or use off-the-shelf algorithms not designed for resource-constrained sensor nodes.

Major attempts have resulted to implement a Sensor-Lempel-Ziv (S-LZW) compression algorithm designed especially for WSNs. While developing Sensor LZW and some simple, but effective variations to this algorithm, Christopher Sadler and Margaret Martonosi in their paper [15] showed how different amounts of compression can lead to energy savings on both the compressing node and throughout the network and that the savings depends heavily on the radio hardware. They achieved significant energy improvements by devising computationally-efficient lossless compression algorithms on the source node. These reduce the amount of data that must be passed through the network, and thus have energy benefits that are multiplicative with the number of hops the data travels through the network. Their evaluation showed reduction in amount of transmitting data by a factor of 4.5.

Regarding this trend for designing efficient data compression algorithms in WSNs, R.Vidhyapriya1 and P. Vanathi designed and implemented two lossless data compression algorithms integrated with the shortest path routing technique to reduce the raw data size and to accomplish optimal trade-off between rate, energy, and accuracy in a sensor network. [16]

## VIII. Conclusion

Today, with growing amount of data storage and information transmission, data compression techniques have a significant role. Even with the advances in bandwidth and storage capabilities, if data were not compressed, many applications would be too costly and the users could not use them. In this research survey, I attempted to introduce two types of compression, lossless and lossy compression, and some major concepts, algorithms and approaches in data compression and discussed their different applications and the way they work. We also evaluated two of the most important compression algorithms based on simulation results. Then as my next contribution, I thoroughly discussed two major everyday applications regarding data compression; JPEG as an example for image compression and MPEG as an example of video compression in our everyday life. At the end of this survey I discussed major issues in leveraging data compression algorithms and the state-of-the art research works done regarding energy saving in top-world-discussed area in networking which is Wireless Sensor Networks.

## References

[1] D. Huffman, "A method for the construction of minimum-redundancy codes, huffman original paper," *Proceedings of the I.R.E*, 1952.

[2] D. Lelewer and D. Hirschberg, "Data compression," *ACM Computing Surveys*, 1987.

[3] R. Pasco, "Source coding algorithms for fast data compression," Ph.D. dissertation, Stanford University, 1976.

[4] K. Sayood, "Lossless compression handbook," *Academic Press*, 2003.

[5] C. Zeeh, "The lempel ziv algorithm," *Seminar Famous Algorithms*, 2003.

[6] T. Welch, "A technique for high-performance data compression," *IEEE Computer*, 1984.

[7] P. Deutsch., "Deflate compressed data format specification version 1.3," *RFC 1951, http://www.faqs.org/rfcs/rfc1951.htm*, 1996.

[8] G. J. Rissanen, J.J.; Langdon, "Arithmetic coding," *IBM Journal of Research and Development*, 1979.

[9] A. Gersho and R. Gray, "Vector quantization and signal compression," *Kluwer Academic Publishers*, 1991.

[10] JPEG2000, *http://www.jpeg.org/jpeg2000/*, .

[11] G. K. Wallace, "The jpeg still picture compression standard," *ACM digital multimedia systems, vol. 34*, 1991.

[12] *A Guide to Image Processing and Picture Management*. Gower Publishing Limited, 1994.

[13] "Digital video coding standards and their role in video communications," 1995.

[14] A. C. W. Paper, "An explanation of video compression techniques," , 2008.

[15] C. M. Sadler and M. Martonosi, "Data compression algorithms for energy-constrained devices in delay tolerant networks," *Proceeding of ACM SenSys*, 2006.

[16] R. Vidhyapriya1 and P. Vanathi, "Energy efficient data compression in wireless sensor networks," *The International Journal of Information Technology*, 2009.